\documentclass[%
 reprint,
 amsmath,amssymb,
 aps,
 superscriptaddress,
 prd,
]{revtex4-2}

\usepackage[colorlinks,
           bookmarks=true,
           linkcolor=blue,
           urlcolor=blue, 
            anchorcolor=black,
            citecolor=blue
            ]{hyperref}

\usepackage{multirow}
\usepackage{graphicx}% Include figure files
\usepackage{dcolumn}% Align table columns on decimal point
\usepackage{bm}% bold math
\usepackage{float}
\usepackage{exscale}
\usepackage{relsize}

\begin{document}

%\preprint{APS/123-QED}

\title{QCD phase transition at finite temperature and chemical potential with the non-extensive statistics}% Force line breaks with \\

%\author{Chao Zhang}
%\affiliation{Key Laboratory of Quark and Lepton Physics (MOE) and Institute 
%of Particle Physics, Central China Normal University, Wuhan 430079, China}
%\affiliation{Department of Physics, East Carolina University, 
%  Greenville, North Carolina 27858, USA} 
%\author{Liang Zheng}
%\affiliation{School of Mathematics and Physics, China University of
%  Geosciences (Wuhan), Wuhan 430074, China}
%\author{Shusu Shi} 
%\affiliation{Key Laboratory of Quark and Lepton Physics (MOE) and Institute
%of Particle Physics, Central China Normal University, Wuhan 430079, China}
%\author{Zi-Wei Lin}\email{linz@ecu.edu}
%\affiliation{Department of Physics, East Carolina University, 
 % Greenville, North Carolina 27858, USA} 
\author{Zhi-Ying Qin}
\affiliation{School of Physics and Information Technology, Shaanxi Normal University, Xi'an 710119, China}
\affiliation{School of Physics Science and Engineering, Tongji University, Shanghai 200092, China}
\author{Jia-Hao Shi}
\affiliation{School of Physics and Information Technology, Shaanxi Normal University, Xi'an 710119, China}
\author{Jin-Peng Zhang}
\affiliation{School of Physics and Information Technology, Shaanxi Normal University, Xi'an 710119, China}
\author{Jian Cao}
\affiliation{School of Physics and Information Technology, Shaanxi Normal University, Xi'an 710119, China}
\author{Bo Feng}
\affiliation{School of Physics and Information Technology, Shaanxi Normal University, Xi'an 710119, China}
\author{Wen-Chao Zhang}
\email{wenchao.zhang@snnu.edu.cn}
\affiliation{School of Physics and Information Technology, Shaanxi Normal University, Xi'an 710119, China}
\author{Hua Zheng}
\affiliation{School of Physics and Information Technology, Shaanxi Normal University, Xi'an 710119, China}
\author{Shi-Jun Mao}
\affiliation{School of Physics, Xi’an Jiaotong University, Xi’an, Shaanxi 710049, China}
\date{\today}% It is always \today, today,
             %  but any date may be explicitly specified

\begin{abstract}

The intrinsic fluctuations, memory effects and long-range color interactions in high energy nuclear collisions imply the presence of non-Markovian processes in the fireball evolution, which affects  the thermalization process towards equilibrium and produces a non-extensive behavior. In order to investigate the non-equilibrium  effect on the quantum chromodynamics (QCD) phase transition at finite temperature ($T$) and chemical potential ($\mu$), we apply a non-extensive correction to the equation of state  in the parton (hadron resonance)  gas at high (low) temperature and interpolate these two equation of states with a smooth crossover. The non-extensive statistics is characterized by a non-extensivity parameter $q$, which measures the degrees of deviation from the thermal equilibrium. It is found that the dimensionless thermodynamic quantities such as the entropy density, the pressure, the energy density, the specific heat at constant volume and  the trace anomaly  are sensitive to the deviation of $q$ from unity and they become large both in the hadronic and quark-gluon plasma phases with the increase of $q$. Moreover, this deviation leads to nontrivial corrections of the squared speed of sound ($(c_s^2)_q$) in the vicinity of the critical point ($T_c$) and at lower temperatures. Additionally, these thermodynamic quantities are sensitive to the deviation of $\mu$ from zero. With increasing $\mu$, they become enhanced in both phases. Specifically, for $(c_s^2)_q$,  the value increases near $T_c$ but decreases at lower temperatures. Finally, we observe that our results with $q=1$ agree well with those from the Lattice QCD, the hadron resonance gas model, and the Thermal-Fist fit to the hadron yields in high energy nuclear collisions in the low temperature region up to $T\sim 150$ MeV.

%We explore the QCD phase transition with $\mu=0$ MeV in the chiral and non-chiral limit cases, as well as the transition with finite $\mu$ in the non-chiral limit case. 
%A non-extensive version of equation of state is constructed to investigate the properties of quantum chromodynamics (QCD) phase transition at finite temperature ($T$) and chemical potential ($\mu$).

%, a non-extensive version of EoS is constructed to investigate the properties of quantum chromodynamics (QCD) phase transition at finite temperature ($T$) and chemical potential ($\mu$). It is based on
\end{abstract}

\maketitle

\section{\label{sec:intro}Introduction}
Quantum-Chromodynamics (QCD) predicts that at high temperatures and energy densities there exists a deconfining transition of the matter governed by strong interactions to a new state denoted as the quark-gluon plasma (QGP). QGP is expected to be produced in ultra-relativistic heavy-ion collisions. Exploring the properties of QCD phase transition is one of the main goals in heavy-ion collisions. 

Several effective models, such as the bag model \cite{bag_model}, the Nambu-Jona-Lasinio (NJL) model \cite{NJL_model_1, NJL_model_2}, the linear sigma model \cite{sigma_model}, the quark-meson model \cite{quark_meson_model_1, quark_meson_model_2}, the Dyson-Schwinger equations approach \cite{Ds_equation_1,Ds_equation_2}, the functional renormalization group approach \cite{FRG_1,FRG_2, FRG_3}, the hadron resonance gas (HRG) and statistical models \cite{HRG_1,HRG_2,HRG_3,HRG_4},  and the Lattice QCD (LQCD) \cite{HRG_3,HRG_4,lattice_1, lattice_2}, have been utilized to investigate the properties of the QCD phase transition. In these models, a  thermal equilibrium is assumed and the extensive statistical approach, such as the Boltzmann-Gibbs (BG) statistics, the Bose-Einstein statistics or the Fermi-Dirac statistics is applied.

However, such an approach is only valid when the heat bath of the system is homogeneous. In reality, this condition cannot always be fulfilled. For example, in high-energy nucleus-nucleus (AA) collisions, extreme conditions of density and temperature give rise to strong intrinsic fluctuations, memory effects and long-range correlations \cite{temp_fluct,temp_fluct_1,non_ext_hydro,q_cent_dependence_1}. These conditions imply the presence of non-Markovian processes in the fireball evolution, which affects the thermalization process towards equilibrium and the standard equilibrium distribution \cite{Markovian_effect}. Thus, instead of an equilibrium state, some kind of stationary state near the equilibrium ($q$-equilibrium) is expected to be formed \cite{stationary_state_1, stationary_state_2}.  This phenomenon can be dealt with non-extensive statistics (Tsallis statistics) \cite{Tsallis_statistics} where the degree of deviation from the equilibrium is described by a single parameter $q$. The non-extensivity in a QCD system can also be understood as a result of the fractal structure of the hot and dense system formed in high energy nuclear collisions \cite{fractal_1,fractal_2}.

In recent years, Tsallis statistics has been widely employed to describe the multi-particle production processes in high-energy proton-proton (pp) \cite{nex17, nex18, Tsallis_10, Tsallis_11, Tsallis_12, Tsallis_13, Tsallis_14, Tsallis_15, Tsallis_16} and AA \cite{Tsallis_1, Tsallis_2, Tsallis_3, Tsallis_4, Tsallis_5, Tsallis_6, Tsallis_7, Tsallis_8, Tsallis_9} collisions. In most of these applications, the Tsallis distribution or the Tsallis-extended blast-wave model is simply fitted to the particle transverse momentum ($p_{\rm T}$) spectra. The extracted values of $q$ from the fits are usually larger than 1. The deviation of $q$ from unity represents the intrinsic fluctuations of the temperature \cite{temp_fluct} or of the mean value of the charged-particle multiplicity \cite{mult_fluct} in the hadronizing system. This deviation can be also be related to the fractal dimension of the hadronizing system  \cite{fractal_1,fractal_2}.
In case of high-energy AA collisions, the fluid dynamics is not considered in these applications, which could affect the accuracy of the results. In light of such a situation, (0+1)-, (1+1)-, (2+1)-, and (3+1)-dimensional hydrodynamic models based on non-extensive statistics were developed as a description of relativistic heavy-ion collisions \cite{q_hydro_model_1,q_hydro_model_2,q_hydro_model_3,q_hydro_model_4}. This is a reasonable extension of the hydrodynamic model, as far-from-equilibrium evolution can still obey the hydrodynamic equations of motion, provided that the system is locally isotropized in the co-moving frame of the fluid \cite{q_hydro_1, q_hydro_2}.  It was shown that the non-extensive hydrodynamic model can describe the charged hadron $p_{\rm T}$ spectra up to 6-8 GeV/c, which is larger than the typical validity range of the extensive hydrodynamic description, 2-3 GeV/c. Moreover, the elliptic flow of charged particles was reduced with the introduction of Tsallis statistics to the hydrodynamic model. The Tsallis statistics has also been applied on the nuclear equation of state (EoS) and the QGP EoS in the framework of a relativistic mean-field approach and the bag model, respectively. In the former model, the interactions of nucleons are mediated by the exchange of isoscalar and isovector mesons, $\sigma$, $\omega$, and $\rho$ \cite{nuclear_EOS1, nuclear_EOS2, nuclear_EOS3, nuclear_EOS4}. In the latter model, the quark matter is described as a gas of free quarks and the non-perturbative effects are simulated by the bag constant $B$ \cite{nuclear_EOS1, nuclear_EOS2,  nuclear_EOS4}. It was observed that both the nuclear EoS and the QGP EoS became stiffer when the non-extensive statistic effect was considered. Furthermore, the Tsallis statistics has been taken into account in the NJL model  to describe the QCD phase transition \cite{q_NJL1, q_NJL2, q_NJL3, q_NJL4}. It was found that with the increase of $q$ the critical end point moved toward the direction of larger chemical potential and lower temperature.

In this study, in order to investigate the non-extensive effect on the QCD phase transition at finite temperature and chemical potential, we apply a non-extensive correction to the EoS in the parton (hadron resonance)  gas \cite{HRG_11, HRG_2, IPG} at high (low) temperature and interpolate these two equation of states with a smooth crossover.  We first explore the QCD phase transition with zero chemical potential. In the QGP phase, the dominant excitations are the massive $u$, $d$, and $s$ quarks, while in the hadronic phase they are the hadrons and resonances composed of $u$, $d$, and $s$ quarks with masses below $2.5$ GeV$/c^2$. Then we investigate the QCD phase transition with finite chemical potential. We will investigate the effects of the non-extensive statistics as well as the chemical potential on the temperature dependence of the dimensionless thermodynamic quantities such as the entropy density,  the pressure, the energy density, the trace anomaly, the heat capacity at the constant volume, and the squared speed of sound. Moreover, we will compare our results with those from the HRG, the LQCD, and the fit of the experimental yields of hadrons at the LHC and RHIC energies using the Thermal-Fist package \cite{Thermal_Fist_1}. 

This paper is organized as follows. In sect. \ref{sec:calculation}, we will describe the Tsallis statistics and construct a non-extensive version of EoS by applying the non-extensive correction to the EoS in the parton (hadron resonance)  gas at high (low) temperature and  making a smooth interpolation of these two equation of states. In sect. \ref{sec:results}, we will discuss  the effects of the non-extensive statistics and the chemical potential  on the thermodynamic quantities  and compare our results with different models. Finally, the conclusion is given in sect. \ref{sec:conclusions}.

\section{\label{sec:calculation}The model}
\subsection{Tsallis statistics}
The non-extensive statistics was first proposed by Tsallis \cite{nex1}. It generalizes the BG statistics and defines the entropy of a system as
\begin{equation}\label{eq:non_extensive_entropy}
	S_q=k\left(1-\sum \limits_{\substack{R=1}}^{W} p_R^q\right)/(q-1),
\end{equation}
where $k$ is a positive constant; $W\in N$ is the total number of microscopic configurations of the system; $p_R$ is the probability associated with the $R^{th}$ state and satisfies $\sum_{R=1}^{W}p_R=1$; $q$ characterizes the degree of non-extensivity reflected in the following pseudo-additive rule for a composite system consisting of two individual systems $A$ and $B$,
\begin{equation}\label{eq:non_extensive_entropy_pseudo_additive_rule}
	\frac{S_q(A+B)}{k}=\frac{S_q(A)}{k}+\frac{S_q(B)}{k}+(1-q)\frac{S_q(A)}{k}\frac{S_q(B)}{k}.
\end{equation}
In the limit $q\rightarrow 1$, the expression in Eq. (\ref{eq:non_extensive_entropy}) approaches the BG entropy,
\begin{equation}\label{eq:non_extensive_entropy_q_to_1}
	\lim \limits_{\substack q\rightarrow 1} S_q=-k\sum_{R=1}^{W}p_R \textrm{ln} p_R.
\end{equation}

In order to determine the probability distribution of a state $R$ in a grand canonical system, we need to maximize the entropy under the internal energy and particle number constraints, 
 \begin{eqnarray}
 	\bar{E}=\sum \limits_{\substack{R=1}}^{W} p_R^{q} E_{R} \label{eq:constrain_entropy_1}, \\ 
	\bar{N}=\sum \limits_{\substack{R=1}}^{W} p_R^{q} N_{R}  \label{eq:constrain_entropy_2} ,
 \end{eqnarray}
where $\bar{E}$ and $\bar{N}$ are, respectively, the average energy and the average number of particles. The maximization of the entropy under constraints in Eqs. (\ref{eq:constrain_entropy_1}) and (\ref{eq:constrain_entropy_2}) leads to the variational equation,
\begin{equation}\label{eq:non_extensive_variation}
\small
	\frac{\delta}{\delta p_R}\left(\frac{S_q}{k}-\alpha\sum_{R=1}^{W} p_R - \beta \sum_{R=1}^{W} p_{R}^{q}E_R - \gamma \sum_{R=1}^{W} p_{R}^{q}N_R\right)=0,
\end{equation}
where $\alpha$, $\beta$ and $\gamma$ are the Lagrange multipliers associated with the total probability, the total energy, and the  total number of particles, respectively. Following the standard thermodynamic definitions, such as $\beta=1/T$ and $\gamma=-\beta \mu$ with $T$ and $\mu$, respectively, being the temperature and the chemical potential of the system, the probability distribution of the state $R$ in the grand-canonical ensemble is written as
\begin{equation}\label{eq:non_extensive_p_i}
p_R=[1+(q-1)(E_R-\mu N_R)/T]^{1/(1-q)}/Z_q,
\end{equation}
where 
\begin{equation}\label{eq:non_extensive_z_q}
Z_q=\sum_{R}[1+(q-1)(E_R-\mu N_R)/T]^{1/(1-q)}
\end{equation}
is the partition function of the grand-canonical system.

The quantum state of the system is uniquely determined when occupation numbers of the one-particle
states are provided. The total energy and the total particle numbers of the system in the state $R$ then are automatically given by 
\begin{align}
E_R &=n_1\varepsilon_1+n_2\varepsilon_2+\cdots+n_k\varepsilon_k+\cdots,  \label{eq:non_extensive_E_i}\\
N_R &=n_1+n_2+\cdots+n_k+\cdots\label{eq:non_extensive_N_i}, 
\end{align}
where  $\varepsilon_k$ and $n_k$ are, respectively,  the  energy of a particle and the number of particles in the state $k$. Substituting Eqs. (\ref{eq:non_extensive_E_i}) and (\ref{eq:non_extensive_N_i}) into Eq. (\ref{eq:non_extensive_z_q}),  one gets the  partition function as \cite{derivation_partition_function}
\begin{equation}\label{eq:non_extensive_z_q_wo_correlation}
\begin{aligned}
Z_q&=\sum_{n_1,n_2\cdots,n_k,\cdots}\exp_q\left(-\frac{(\varepsilon_1-\mu)n_1}{T}\right)\\
&\otimes_q \exp_q\left(-\frac{(\varepsilon_2-\mu)n_2}{T}\right)\otimes_q \cdots\\
&\otimes_q \exp_q\left(-\frac{(\varepsilon_k-\mu)n_k}{T}\right)\otimes_q \cdots\\
&=(\prod_{k=1}^{\infty})_q \sum_{n_{k=0}}^{\infty} \exp_q\left(-\frac{(\varepsilon_k-\mu)n_k}{T}\right),\\
\end{aligned}
\end{equation}
where $\textrm{exp}_q(x)=[1+(1-q)x]^{(1/(1-q))}$, $\textrm{exp}_q(x)\otimes_q \textrm{exp}_q(y)=\textrm{exp}_q(x+y)$, $(\prod)_q$ means the successive $q$-multiplication. When taking into account the $d$ internal degrees of freedom for the non-interacting fermions and bosons, the grand-canonical partition function, respectively, becomes
\begin{equation}\label{eq:non_extensive_z_q_dof_fermion}
Z_q=(\prod_{k=1}^{\infty})_q\left[1+{\rm exp}_q\left(-\frac{\varepsilon_k-\mu}{T}\right)\right]^{+d}
\end{equation}
and 
\begin{equation}\label{eq:non_extensive_z_q_dof_boson}
Z_q\approx(\prod_{k=1}^{\infty})_q\left[1-{\rm exp}_q\left(-\frac{\varepsilon_k-\mu}{T}\right)\right]^{-d}.
\end{equation}
The derivations of Eqs. (\ref{eq:non_extensive_z_q_dof_fermion}) and (\ref{eq:non_extensive_z_q_dof_boson}) are shown in the appendix A. In the derivation of Eq. (\ref{eq:non_extensive_z_q_dof_boson}), we have made an approximation that the higher orders $\mathcal{O}(q-1)^2$ have been ignored.

\subsection{Equation of states in the hadronic and QGP phases }\label{sect:EoS}
In the framework of extensive statistics, the grand-canonical potential $\Omega$ is related to the grand-canonical partition function $Z$ as follows \cite{Landau_stat} 
\begin{equation}
\begin{aligned}
\Omega=-T\ln Z.
\end{aligned}
\label{eq9}
\end{equation}
In the non-extensive statistics, the relation between the grand-canonical potential $\Omega_q$ and the grand-canonical partition function $Z_q$ is similar to that in the extensive statistics, 
\begin{equation}
\begin{aligned}
\Omega_q=-T\ln_qZ_q,
\end{aligned}
\label{eq:grand_canonical_potential_nonextensive}
\end{equation}
where the $q$-logarithm ${\rm ln}_q x$ is defined as \cite{nex2}
\begin{equation}
\ln_qx\equiv\frac{x^{1-q}-1}{1-q}.
\label{eq:ln_q}
\end{equation}
The derivation of Eq. (\ref{eq:grand_canonical_potential_nonextensive}) is presented in the appendix B.

It has been shown that in the non-extensive statistics the thermodynamic consistency is also satisfied \cite{nex18,non_ext_hydro}. Thus, the pressure is given by  
\begin{equation}\label{eq:nonextensive_pressure_1}
\begin{aligned}
P_q&=-\frac{\Omega_q}{V}\\
&=\frac{T\ln_qZ_q}{V}\\
&=\frac{T}{V}\sum_{k=1}^{\infty} \ln_q \left[1\pm{\rm exp}_q\left(-\frac{\varepsilon_k-\mu}{T}\right)\right]^{\pm d},\\
\end{aligned}
\end{equation}
where the identity $\ln_q (x\otimes_q y)=\ln_q x + \ln_q y$ is used.
%\begin{equation}
%s_q=\frac{\partial P_q}{\partial T}.
%\label{eq:nonextensive_entropy}
%ß\end{equation}
In the large volume limit, the summation in Eq. (\ref{eq:nonextensive_pressure_1}) is replaced by the integral in the 6-dimensional phase space,
\begin{equation}
\sum_k\rightarrow V\int\frac{d^3k}{(2\pi)^3}.
\label{eq14}
\end{equation}
Therefore, the pressure is rewritten as
\begin{equation}\label{eq:nonextensive_pressure}
P_q= T\int\frac{d^3k}{(2\pi)^3}\ln_q \left[1\pm{\rm exp}_q\left(-\frac{\varepsilon_k-\mu}{T}\right)\right]^{\pm d},
\end{equation}
where $\varepsilon_k=\sqrt{m^2+k^2}$, $m$ and $k$ are, respectively, the rest mass and the momentum of the particle. 

The entropy density for a multi-component system with $N$ number of species then is expressed as
\begin{equation}\label{eq:nonextensive_entropy}
s_q= \sum_{i=1}^N\partial P_q^i/\partial T,
\end{equation}
where $P_q^i$ is the pressure for the single-component system with the particle type  $i$. One key factor in the calculation of the entropy density is  the internal degrees of freedom  $d$.  In the QGP (hadronic) phase, the masses of $u$, $d$, and $s$ quarks (hadrons and resonances composed of these quarks) are considered. The degeneracy factor for quarks (antiquarks) of a given flavor is $d_{q(\bar{q})} = 2_{\text{spin}} \times N_c = 6$, while for gluons the degeneracy is $d_g = 2_{\text{spin}} \times (N_c^2 - 1) = 16$, where $N_c = 3$ represents the number of colors. For hadrons of a specific species, the degeneracy factor is given by \(d_H = 2S + 1\), with \(S\) denoting the spin of the hadron. Another key factor in the calculation of the entropy density is the chemical potential. The chemical potential for the $i^{th}$ particle species is $\mu_i=B_i\mu_B+S_i\mu_S+Q_i\mu_Q$, where $B_i$, $S_i$, and $Q_i$ are, respectively, the particle's baryon, strangeness, and electric charge numbers, $\mu_B$, $\mu_S$, and $\mu_Q$ are the respective chemical potentials of the system.

\subsection{Smooth crossover around the critical temperature}
In section \ref{sect:EoS}, the EoS in the hadronic (QGP) phase has been constructed by applying a non-extensive correction to the hadron resonance (parton) gas model.  In order to match the equation of states in the hadronic and QGP phases around the crossover temperature $T_c$, we make a smooth interpolation of $s_q(T)$ between the hadronic gas at low $T$ and the QGP at high $T$. A simplest possible parametrization of the entropy density $s_q(T)$ satisfying the  thermodynamic inequality $\partial s_q(T)/\partial T\geq 0$ and the third law of thermodynamics $s_q(T\rightarrow0)\rightarrow \textrm{constant}$ is \cite{HRG_1}
\begin{equation}
s_q(T)=f(T)s_q^{\textrm{H} }(T)+(1-f(T))s_q^{\textrm{QGP} }(T),
\label{eq:entropy_mixed_phase}
\end{equation}
where $f(T)$ is a weighting function,
\begin{equation}
f(T)=\frac{1}{2}[1-{\rm tanh}((T-T_c)/\it \Gamma)],
\label{eq:weight_function}
\end{equation}
with $\it \Gamma$ setting the width of the phase transition region. For $T$ satisfying $\lvert T-T_c\rvert > \it \Gamma$, $s_q(T)$ quickly approaches $s_q^{\textrm{H}}$ below $T_c$ and $s_q^{\textrm{QGP}}$ above $T_c$.

The pressure $P_q(T)$, the energy density $\varepsilon_q(T)$, the specific heat at constant volume $(C_V)_q(T)$, and the squared speed of sound  $(c_s^2)_q(T)$  then can be, respectively, obtained using the following expressions,
\begin{equation}
P_q(T)=\int_0^Ts_q(t)dt,
\label{eq:pressure_mixed_phase}
\end{equation}
\begin{equation}
\varepsilon_q(T)=Ts_q(T)-P_q(T)+\sum_i \mu_i\int_0^T \frac{\partial s_q(t)}{\partial \mu_i} dt,
\label{eq19}
\end{equation}
\begin{equation}
\begin{aligned}
(C_V)_q(T)=\frac{\partial \varepsilon_q(T)}{\partial T}\big |_V=T\frac{\partial s_q(T)}{\partial T}+\sum_i \mu_i\frac{\partial s_q(T)}{\partial \mu_i},
\end{aligned}
\label{eq20}
\end{equation}
\begin{equation}
\begin{aligned}
(c_s^2)_q(T)=\frac{\partial P_q(T)}{\partial\varepsilon_q(T)}=\frac{s_q(T)}{(C_V)_q(T)},
\end{aligned}
\label{eq21}
\end{equation}
where the index $i$ in the summation runs over all the particle species.

\section{\label{sec:results}Results and discussions}
\subsection{Zero chemical potential}
The QCD phase transition is first studied at zero chemical potential. In the QGP phase, the $u$, $d$, and $s$ quarks are considered, with masses set to $m_{u}=3.503\text{ MeV}/c^2$, $m_{d}=3.503\text{ MeV}/c^2$, and $m_{s}=96.4\text{ MeV}/c^2$, respectively \cite{mass_quark}. In the hadronic phase, hadrons and hadron resonances composed of $u$, $d$, and $s$ quarks with masses below $2.5$ GeV$/c^2$ \cite{lattice_1} from the Particle Data Group \cite{pdg_2016} are considered.

\begin{figure*}[]
\begin{center}
		\includegraphics[scale=1.1]{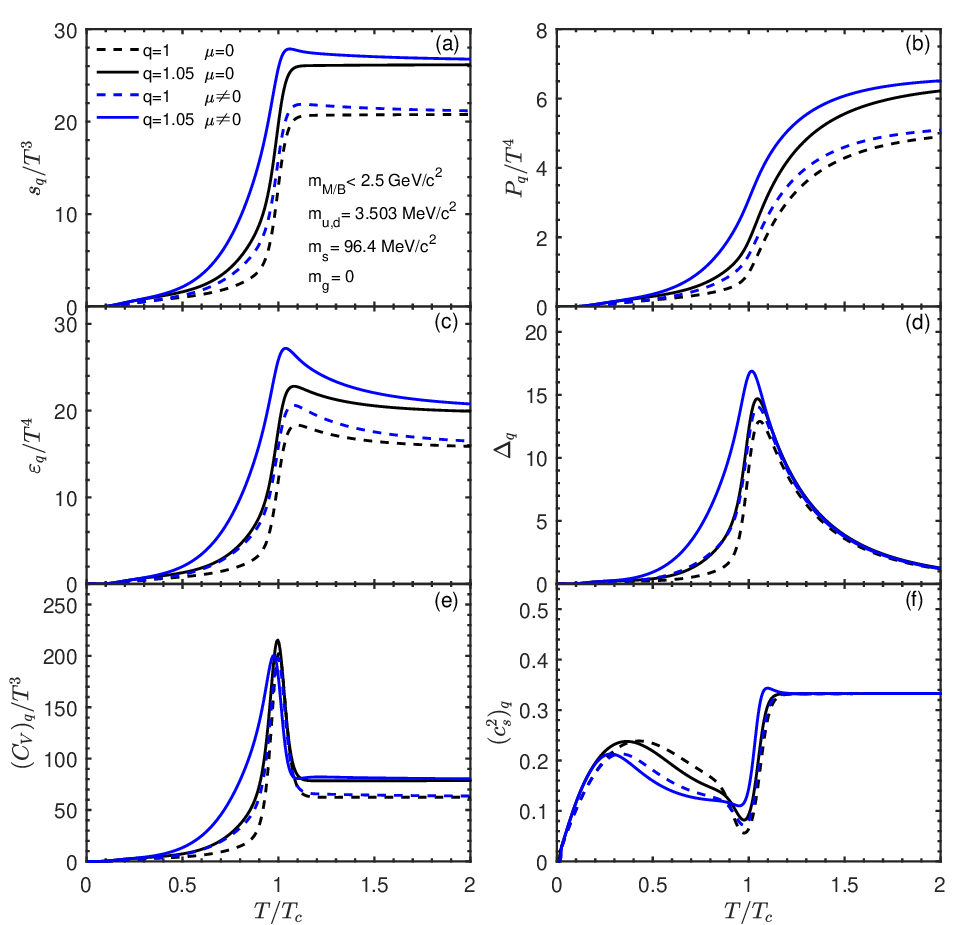}
		\caption{(a) ((b), (c), (d), (e), and (f)): $s_q/T^3$ ($P_q/T^4$, $\varepsilon_q/T^4$, $\Delta_q$, $(C_V)_q/T^3$, and $(c_s^2)_q$) as a function of $T/T_c$.  The solid and dashed black  lines correspond to $q=$ 1.05 and $q=$ 1 with zero chemical potential, respectively. The solid and dashed blue lines correspond to $q=$ 1.05 and $q=$ 1 with $\mu_B=$ 398.2 MeV, $\mu_S=$ 89.5 MeV, and $\mu_Q=$ 0 MeV, respectively.}
		\label{fig:thermodynamic_quantities_mu}
\end{center}
\end{figure*}

In each subplot of Fig. \ref{fig:thermodynamic_quantities_mu}, the dashed and solid back curves, respectively, present the dimensionless thermodynamic quantities such as the entropy density $s_q/T^3$, the pressure $P_q/T^4$, the energy density $\varepsilon_q/T^4$, the specific heat at constant volume $(C_V)_q/T^3$, the trace anomaly $\Delta_q=(\varepsilon_q-3P_q)/T^4$, and the squared speed of sound $(c_s^2)_q$ as a function of the temperature for two non-extensive parameters $q=$ 1 and $q=$ 1.05.  In the figure, $T_c$ is taken to be 156.5 MeV \cite{T_c}, $\it \Gamma$ is set as 0.05 $T_c$, which is identical to the choice made in Ref. \cite{HRG_1}. The qualitative behavior of these thermodynamic quantities in both cases can be summarized as follows. 

(i) There is a rapid increase of the entropy density in a narrow region of $T$ around $T_c$ (with a width $\sim$ 8 MeV). In the high temperature region, $s_q/T^3$ approaches the Stefan-Boltzmann (SB) limit. In the low temperature region, $s_q/T^3$ approaches 0. This behavior could be explained as follows. The excitation of a massive hadron of mass $m$ is exponentially suppressed by the Boltzmann factor $\exp(-m/T)$. The entropy density $s_q$ is dominated by this exponential factor, which vanishes much more rapidly than the power-law term $T^3$ when $T\rightarrow 0$.

%, which is consistent with the the Nernst’s theorem \cite{Landau_stat}. 

(ii) $P_q/T^4$ increases rather slowly with $T>T_c$ and approaches the SB limit at high temperature. As shown in Eq. (\ref{eq:pressure_mixed_phase}), $P_q(T)$ is expressed as an integral of $s_q(T)$, thus such a continuous and slow rise is naturally expected \cite{HRG_1}. 

(iii) There is a peak just above $T_c$ for the temperature dependence of $\varepsilon_q/T^4$.  The existence of such a peak is due to the rapid increase of $s_q(T)$ and the slow rise of $P_q(T)$ above $T_c$ \cite{HRG_1}.

(iv) $(C_V)_q/T^3$ has an obvious peak near $T_c$. The specific heat at constant volume is defined as the derivative of the energy density with respect to the temperature. The dimensionless $\varepsilon_q/T^4$ has a peak above  $T_c$, which leads to the peak of $(C_V)_q/T^3$ near $T_c$.

(v) $\Delta_q$ also has a prominent peak near $T_c$. Such a peak is a consequence of the rapid increase of $s_q(T)$. It vanishes when the temperature approaches either0 or $\infty$.

(vi) $s_q/T^3$, $P_q/T^4$, $\varepsilon_q/T^4$, $(C_V)_q/T^3$, and  $\Delta_q$ are sensitive to the deviation of $q$ from unity and they become large both in the hadronic and QGP phases with the increase of $q$. This can be interpreted as the effect that at a fixed temperature the particle yield in the non-extensive distribution with $q>$ 1 is always larger than that in the extensive one. %Moreover, as the degrees of freedom is larger in the hadronic phase than in the QGP phase, the increase is more obvious in the hadronic phase than in the QGP phase. 

\begin{figure*}[t]
\begin{center}
		\includegraphics[scale=1.1]{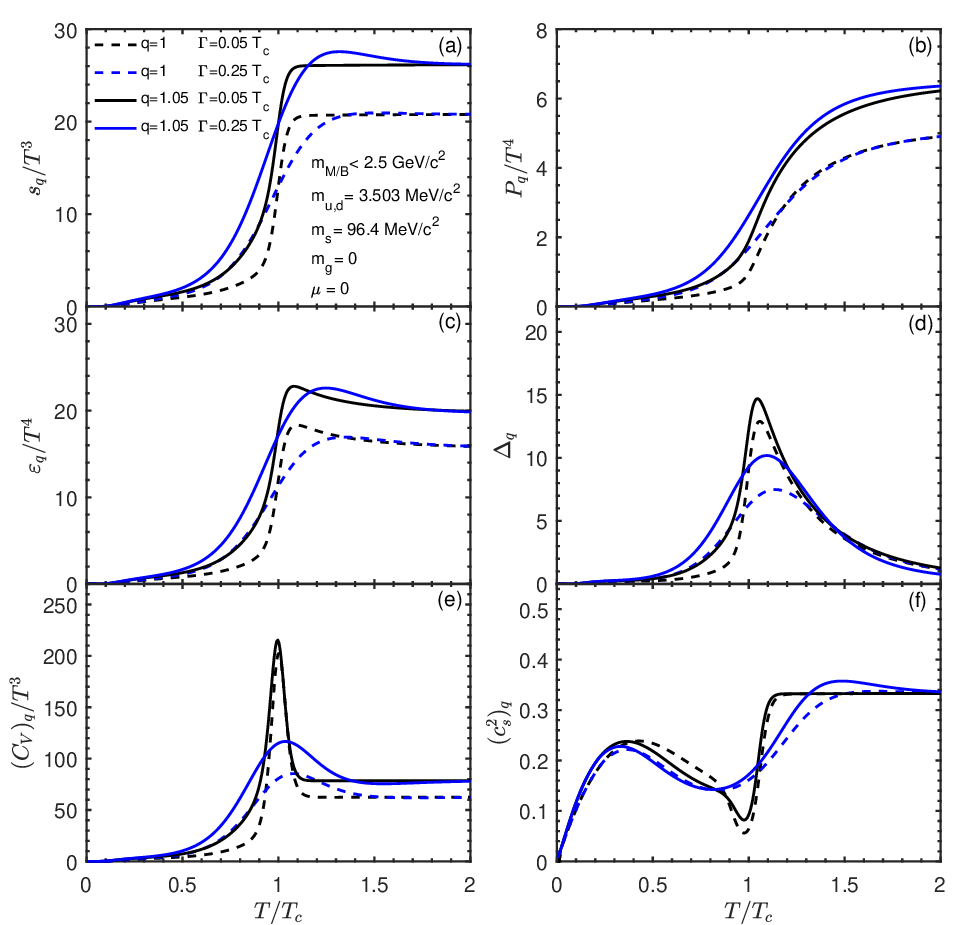}
		\caption{(a) ((b), (c), (d), (e), and (f)): $s_q/T^3$ ($P_q/T^4$, $\varepsilon_q/T^4$, $\Delta_q$, $(C_V)_q/T^3$, and $(c_s^2)_q$) as a function of $T/T_c$ at zero chemical potential.  The dashed (solid) black and blue curves, respectively, represents the temperature dependence of these quantities with $\it \Gamma=\textrm{0.05 }T_c$ and  $\it \Gamma=\textrm{0.25 } T_c$ for the extensive (non-extensive) case.}
		\label{fig:thermal_dependence_gamma}
\end{center}
\end{figure*}

(vii) $(c_s^2)_q$ exhibits a sharp drop in the critical region $|T - T_c| <\it \Gamma$ and stays below 1/3 in the hadronic phase.  In both cases, the behavior results from the energy density $\varepsilon_q(T)$ rising more rapidly than the pressure $P_q(T)$, leading to a soft equation of state. Across $T_c$, this is due to deconfinement. In the hadronic phase, it is a consequence of hadron masses. Moreover, $(c_s^2)_q$ approaches $1/3$ in the high temperature region and 0 in the low temperature region. This could be understood as follows. At low temperatures, the exponential suppression $\exp(-m/T)$ causes both $P_q$ and $\varepsilon_q$ to vanish, but their derivatives vanish at different rates. The ratio of these derivatives, $(c_s^2)_q=\partial P_q/\partial \varepsilon_q$, scales as $T/m$, guaranteeing it approaches zero with $T\rightarrow 0$. At extremely high temperatures, the masses of the $u$, $d$, and $s$ are negligible. Thus, these quarks behave dynamically as massless non-interacting particles.
Additionally, the deviation of $q$ from unity leads to nontrivial corrections of the squared speed of sound in the vicinity of the critical point and at lower temperatures. 
%Moreover, in the former case, $(c_s^2)_q$ is insensitive to the deviation of $q$ from unity. However, in the latter case,  this deviation leads to nontrivial corrections of the sound velocity in the vicinity of the critical point and at lower temperatures. 
%There is a sudden drop of $(c_s^2)_q$ in the narrow region $|T-T_c|<\it \Gamma$. This is a direct result of the slow variation of $P_q(T)$ compared to the rapid variation of $\varepsilon_q(T)$ across $T_c$, which softens the equation of state. In the hadronic phase, the energy density $\varepsilon_q(T)$ increasing faster than the pressure $P_q(T)$ will also make $(c_s^2)_q$ less than the massless limit of 1/3.

We have also investigated the $\it \Gamma$ dependence of the thermodynamic
quantities at the vanishing chemical potential for both extensive and non-extensive cases in  Fig. \ref{fig:thermal_dependence_gamma}. The dashed (solid) black and blue curves, respectively, represents the temperature dependence of these quantities with $\it \Gamma=\textrm{0.05 }T_c$ and  $\it \Gamma=\textrm{0.25 } T_c$ for the extensive (non-extensive) case. It is observed that an increase in the value of $\it \Gamma$ leads to a more gradual temperature dependence of the entropy density and pressure. Concurrently, the characteristic peaks in the trace anomaly and the specific heat at constant volume undergo a broadening and a reduction in height. For the squared speed of sound, an increase in $\it \Gamma$ produces a dip that is both shallower and significantly wider. This set of phenomena is attributable to the broadening of the QCD phase transition window. Specifically, a larger $\it \Gamma$ implies that the conversion from  the hadronic phase to the QGP phase takes place across an extended temperature interval, which smears out the thermodynamic profiles, resulting in smoother, broader, and flatter curves.

\subsection{Finite chemical potential}

The QCD phase transition is also investigated at finite chemical potential. The baryon, strangeness, and electric charge potentials  are, respectively, set as $\mu_B=$ 398.2 MeV, $\mu_S=$ 89.5 MeV, and $\mu_Q=$ 0 MeV, which correspond to the chemical potentials of the system created in the central Au-Au collisions at $\sqrt{s_{\rm NN}}=$ 7.7 GeV \cite{chemical_potential}. In each subplot of Fig. \ref{fig:thermodynamic_quantities_mu}, the solid and dashed blue curves, respectively, show the temperature dependence of the thermodynamic
quantities  with finite chemical potential for  $q=$ 1.05 and $q=$ 1. For the case with finite chemical potential, $T_c$  and $\it \Gamma$ are set to be the same values as those for the case with vanishing chemical potential. The qualitative behaviors of the thermodynamic quantities are summarized as follows.

%(i) The temperature dependence of $s_q/T^3$, $P_q/T^4$, $\varepsilon_q/T^4$, $(C_V)_q/T^3$, $\Delta_q$ and $(c_s^2)_q$ with finite chemical potential is similar to that with zero chemical potential.

(i) $s_q/T^3$, $P_q/T^4$, $\varepsilon_q/T^4$, $(C_V)_q/T^3$, and  $\Delta_q$ are sensitive to the deviation of the chemical potential from zero and they become large both in the hadronic and QGP phases with the increase of $\mu$. This is due to the reason that at a given temperature a larger population is expected for the system with finite chemical potential. A similar increase was observed for the normalized pressure and trace anomaly obtained from the Dyson-Schwinger equations when the values of quark chemical potential were increased \cite{QCD_PT_chemical_potential}.

(ii) With the increase of the chemical potential, the squared speed of sound $(c_s^2)_q$ increases in the vicinity of $T_c$ and decreases at lower temperatures, which is different from the chemical potential dependence of $s_q/T^3$, $P_q/T^4$, $\varepsilon_q/T^4$, $(C_V)_q/T^3$ and  $\Delta_q$.

(iii) At high temperature, these thermodynamic quantities with finite chemical potential approach the respective ones with zero chemical potential. At low temperature, the thermodynamic quantities with both zero and finite chemical potential approach 0 when $T\rightarrow 0$.

(iv) As shown in the previous section, at zero chemical potential the thermodynamic quantities are sensitive to the deviation of $q$ from unity. With the consideration of both the effects of finite chemical potential and the non-extensivity, an obvious change of the behavior for the thermodynamic quantities is observed.

\subsection{Comparison with other models}

\begin{figure}[]
\begin{center}
		\includegraphics[scale=0.8]{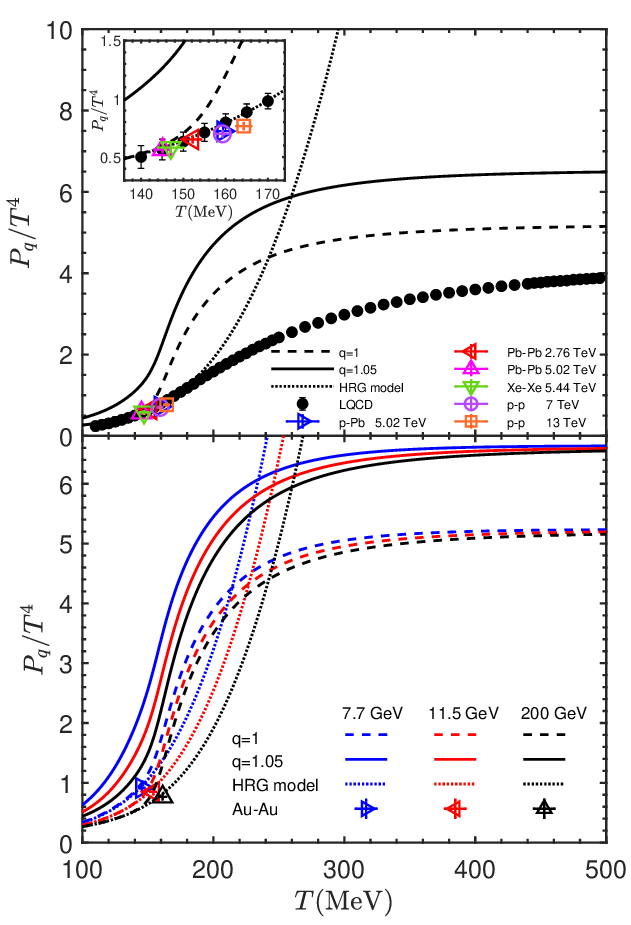}
		\caption{Upper panel: $P_q/T^4$ as a function of $T$ with zero chemical potential for $q=$ 1.05 (the solid line) and 1 (the dashed line). The results from the HRG model, the LQCD,  and the Thermal-Fist fit  are, respectively, shown as the dotted line, the solid black circles, and the empty markers. The inset is the temperature dependence of $P_q/T^4$ in the region with 136 MeV $<T<174$ MeV. Lower panel: $P_q/T^4$ as a function of $T$ with finite chemical potential for $q=$ 1.05 (the solid line) and $q=$ 1 (the dashed line). The results from the HRG model and the Thermal-Fist fit  are, respectively, shown as the dotted line and the empty markers.}
		\label{fig:pT4_HRG_LQCD_Thermalfist}
\end{center}
\end{figure}

The upper panel of Fig. \ref{fig:pT4_HRG_LQCD_Thermalfist} displays the comparison of the temperature dependence of  $P_q/T^4$ calculated with zero chemical potential for different models. Our results with $q=$ 1.05 and $q=$ 1 are, respectively, presented as the solid and dashed lines. The result from the HRG model is shown as the dotted line, while the result from the LQCD is presented as the solid black circles. The dependence  extracted from the fit of experimental yields of hadrons at the central centrality in pp collisions at $\sqrt{s}=$ 7 TeV \cite{pp_yield_1, pp_yield_2, pp_yield_3} and 13 TeV \cite{pp_yield_4, pp_yield_5, pp_yield_6}, p-Pb collisions at $\sqrt{s_{\rm NN}}=$ 5.02 TeV \cite{p_Pb_yield_1, p_Pb_yield_2, p_Pb_yield_3}, Pb-Pb collisions at $\sqrt{s_{\rm NN}}=$ 2.76 TeV \cite{Pb_Pb_yield_1, Pb_Pb_yield_2, Pb_Pb_yield_3, Pb_Pb_yield_4} and 5.02 TeV \cite{Pb_Pb_yield_5, Pb_Pb_yield_6}, and Xe-Xe collisions at $\sqrt{s_{\rm NN}}=$ 5.44 TeV \cite{Xe_Xe_yield_1}  using the Thermal-Fist package \cite{Thermal_Fist_1} are shown as empty markers. Thermal-Fist is a C++ package designed for the analyses of particle productions in relativistic heavy-ion collisions within the framework of the HRG model. In the package, there are several options regarding the ensemble with which to treat the baryon number, strangeness, and electric charge. In this work, we choose the grand canonical ensemble, which is  widely applied to heavy-ion collisions \cite{grand_canonical_1,grand_canonical_2,grand_canonical_3,grand_canonical_4}.  It is shown that our result with $q=1$ agrees well with the HRG model and the LQCD data up to $T\sim$ 150 MeV.  In our model, $P_q(T)$ quickly approaches the pressure at the hadronic phase with $T<T_c-\it \Gamma$. When $q\rightarrow 1$, the Tsallis statistics approaches the BG statistics valid for systems under the thermal equilibrium. Thus, the agreement at low temperature is reasonably expected as the HRG model and the LQCD calculations are based on equilibrium considerations. Moreover, our result with $q=1$ is consistent with the results from the Thermal-Fist fit to the hadron yields at the central centrality in Pb-Pb collisions at $\sqrt{s_{\rm NN}}=$ 2.76 TeV (triangle-left) and 5.02 TeV (triangle-up), and in Xe-Xe collisions at $\sqrt{s_{\rm NN}}=$ 5.44 TeV (triangle-down). However, it obviously overestimates the results from the Thermal-Fist fit to the hadron yields in p-Pb collisions at $\sqrt{s_{\rm NN}}=$ 5.02 TeV (triangle-right), and in pp collisions at $\sqrt{s}=$ 7 TeV (empty circle) and 13 TeV (empty square). In pp and p-Pb collisions, the temperature of the system is around 160 MeV, which is in the phase transition region $\lvert T-T_c\rvert < \it  \Gamma$, leading to the overestimate of our model for the pressure. At high temperature, our result approaches the SB limit of the pressure for idea parton gas. However, in LQCD, partons interact with each other, which leads to the deviation from the limit. In the HRG model, only the hadron gas is considered and the phase transition is absent, thus the pressure approaches infinity at high temperature. 

\begin{table}[]
\caption{Chemical potentials in central Au-Au collisions at $\sqrt{s_{\rm NN}}=$ 7.7, 11.5, and 200 GeV. }\label{tab:Au_Au_chemical_potential}
	\begin{ruledtabular}
	\begin{tabular}{cccccc}
		$\sqrt{s_{\rm NN}}$ (GeV)   & centrality & $\mu_B$ (MeV) & $\mu_S$ (MeV)& $\mu_Q$ (MeV) \\
		\colrule
		7.7  & 0-5$\%$ & 398.2& 89.5&0 \\ 
       11.5 & 0-5$\%$ & 287.3& 64.5&0\\ 
       200 & 0-5$\%$ & 28.4& 5.6& 0\\
	\end{tabular}
    \end{ruledtabular}
\end{table}

The lower panel of Fig. \ref{fig:pT4_HRG_LQCD_Thermalfist} shows the comparison of the temperature dependence of  $P_q/T^4$ calculated with three sets of finite chemical potentials (see table \ref{tab:Au_Au_chemical_potential}) for different models. They correspond to the chemical potentials in central  Au-Au collisions at $\sqrt{s_{\rm NN}}=$ 7.7, 11.5, and 200 GeV \cite{chemical_potential}.  Our results with $q=$ 1.05 and $q=$  1 are, respectively, presented as the solid and dashed lines. The results from the HRG model are shown as the dotted line. The dependence  extracted from the Thermal-Fist fit of experimental yields of hadrons in central Au-Au collisions at $\sqrt{s_{\rm NN}}=$ 7.7, 11.5, and 200 GeV \cite{chemical_potential, Au_Au_1, Au_Au_2, Au_Au_3} are shown as empty blue, red, and black markers, respectively. It is found that  our results with $q=1$ agree well with the HRG model up to $T\sim$ 150 MeV.  Moreover, our results with $q=1$ are consistent with the results from the Thermal-Fist fit to the hadron yields in central Au-Au collisions at $\sqrt{s_{\rm NN}}=$ 7.7 GeV (triangle-right) and 11.5 GeV (triangle-left). However, they overestimate the result from the Thermal-Fist fit to the hadron yields in central Au-Au collisions at $\sqrt{s_{\rm NN}}=$ 200 GeV (triangle-up). The temperature of the system in Au-Au collisions at $\sqrt{s_{\rm NN}}=$ 200 GeV is around 160 MeV, falling in the phase transition region and leading to this overestimate.

\section{Conclusions}\label{sec:conclusions}

In this work, by applying the non-extensive correction to the EoS in the parton (hadron resonance)  gas at high (low) temperature and making a smooth interpolation of these two equation of states, we have constructed a non-extensive version of EoS  to investigate the properties of QCD phase transition with both zero and non-zero chemical potentials. We find that the dimensionless quantities such as $s_q/T^3$, $P_q/T^4$, $\varepsilon_q/T^4$, $(C_V)_q/T^3$, and $\Delta_q$ are highly sensitive to the non-extensivity parameter $q$. As $q$ deviates from unity, these quantities increase in both the hadronic and QGP phases. Furthermore, this deviation introduces nontrivial corrections to $(c_s^2)_q$ near the critical point and at lower temperatures. These thermodynamic quantities also exhibit significant sensitivity to the chemical potential $\mu$. They increase with $\mu$ in both phases.  Specifically, $(c_s^2)_q$ rises near the phase transition temperature but decreases at lower temperatures. Finally, we observe that our results with $q=1$ agree well with the LQCD, the HRG, and the results form the Thermal-Fist fit to the hadron yield in high energy nuclear collisions in the low temperature region up to $T\sim 150$ MeV. These findings will deepen our understanding for the properties of the QCD phase transition at finite temperature and chemical potential both in the extensive and non-extensive statistics.

%We find that the dimensionless quantities $s_q/T^3$, $P_q/T^4$, $\varepsilon_q/T^4$, $(C_V)_q/T^3$ and  $\Delta_q$ are sensitive to the deviation of $q$ from unity and they become large both in the hadronic and QGP phases with the increase of $q$. Moreover, this deviation leads to nontrivial corrections of $(c_s^2)_q$ in the vicinity of the critical point and at lower temperatures. Moreover, these dimensionless thermodynamic quantities are sensitive to the deviation of $\mu$ from zero and they become large both in the hadronic and QGP phases with the increase of $\mu$. However, $(c_s^2)_q$ increases in the vicinity of the phase transition temperature and decreases at lower temperature. Finally, in the non-chiral limit case,  we observe that our results with $q=1$ agree well with the LQCD, the HRG and the results form the Thermal-Fist fit to the hadron yield in high energy nuclear collisions in the low temperature region up to $T\sim 150$ MeV. These findings will deepen our understanding for the properties of the QCD phase transition at finite temperature and chemical potential both in the extensive and non-extensive statistics.

\begin{acknowledgments}

This work is supported by the Scientific Research Foundation for the Returned Overseas Chinese Scholars, State Education Ministry, by Natural Science Basic Research Plan in Shaanxi Province of China (program No. 2023-JC-YB-012), and by the National Natural Science Foundation of China under Grant Nos. 11447024,  11505108 and 12275204.
\end{acknowledgments}

\section*{Appendix A: The derivation of Eqs. (\ref{eq:non_extensive_z_q_dof_fermion}) and (\ref{eq:non_extensive_z_q_dof_boson})}\label{Appendix:A}
\setcounter{equation}{0}
\setcounter{subsection}{0}
\renewcommand{\theequation}{A\arabic{equation}}
In the framework of the non-extensive statistics, the grand-canonical partition function of particles with $d$ internal degrees of freedom is given by
\begin{equation}
\begin{aligned}
Z_q=(\prod_{k=1}^{\infty})_q\left[\sum_{n_k=0}^{\infty}\exp_q\left(-\frac{(\varepsilon_k-\mu)n_k}{T}\right)\right]^d,
\label{eqA1}
\end{aligned}
\end{equation}
where $\varepsilon_k$ and $n_k$ are, respectively, the energy of a particle and the number of particles in the state $k$; $T$ and $\mu$ are, respectively, the temperature and the chemical potential of the system. 

For non-interacting fermions, their grand-canonical partition function is written as 
\begin{equation}
\begin{aligned}
Z_q^{F}&=(\prod_{k=1}^{\infty})_q\left[\sum_{n_k=0}^{1}\exp_q\left(-\frac{(\varepsilon_k-\mu)n_k}{T}\right)\right]^d\\
&=(\prod_{k=1}^{\infty})_q\left[1+\exp_q\left(-\frac{(\varepsilon_k-\mu)}{T}\right)\right]^d.
\label{eqA2}
\end{aligned}
\end{equation}
For non-interacting bosons, their grand-canonical partition function is written as
\begin{equation}
\begin{aligned}
Z_q^{B}=&(\prod_{k=1}^{\infty})_q\left[\sum_{n_k=0}^{\infty}\exp_q\left(-\frac{(\varepsilon_k-\mu)n_k}{T}\right)\right]^d\\
=&(\prod_{k=1}^{\infty})_q\left[1+\cdots+{\rm exp}_q\left(-n_k\frac{\varepsilon_k-\mu}{T}\right)+\cdots\right]^d\\
\approx&(\prod_{k=1}^{\infty})_q\left[1-{\rm exp}_q\left(-\frac{\varepsilon_k-\mu}{T}\right)\right]^{-d}, 
\label{eqA3}
\end{aligned}
\end{equation}
where  the approximation  $\textrm{exp}_q(x)\textrm{exp}_q(y)=\textrm{exp}_q(x+y+(1-q)xy)\approx\textrm{exp}_q(x+y)$ has been adopted \cite{Tsallis_stat}.

\section*{Appendix B: The derivation of Eq. (\ref{eq:grand_canonical_potential_nonextensive})}\label{Appendix:B}
\setcounter{equation}{0}
\setcounter{subsection}{0}
\renewcommand{\theequation}{B\arabic{equation}}
In the grand-canonical ensemble, the probability distribution of the state $R$ and the partition function  in Eqs. (\ref{eq:non_extensive_p_i}) and (\ref{eq:non_extensive_z_q}) are, respectively, rewritten as
\begin{equation}
\begin{aligned}
p_R=\frac{1}{Z_q}\exp_q(-\alpha N_R-\beta E_R),\\
Z_R=\sum_{R=1}^W\exp_q(-\alpha N_R-\beta E_R),
\end{aligned}
\label{eqA6}
\end{equation}
where $\alpha\equiv-\mu/T$, $\beta\equiv1/T$, $W$ is the total number of states.

The particle number $N$, the energy $E$, and the generalized force $Y$ are then, respectively, given by
\begin{equation}
\begin{aligned}
N&=\sum_R p_R^q N_R=\frac{\sum_R[\exp_q(-\alpha N_R-\beta E_R)]^qN_R}{Z_q^q}\\
&=-\frac{\partial}{\partial\alpha}\ln_qZ_q,
\end{aligned}
\label{eq:N}
\end{equation}
\begin{equation}
\begin{aligned}
E&=\sum_ip_R^qE_R=\frac{\sum_R[\exp_q(-\alpha N_R-\beta E_R)]^qE_R}{Z_q^q}\\
&=-\frac{\partial}{\partial\beta}\ln_qZ_q,
\end{aligned}
\label{eq:E}
\end{equation}
\begin{equation}
\begin{aligned}
Y&=\sum_Rp_R^q\frac{\partial E_R}{\partial y}=\frac{\sum_R[\exp_q(-\alpha N_R-\beta E_R)]^q\frac{\partial E_R}{\partial y}}{Z_q^q}\\
&=-\frac{1}{\beta}\frac{\partial}{\partial y}\ln_qZ_q.
\end{aligned}
\label{eq:Y}
\end{equation}

With Eqs. (\ref{eq:N}), (\ref{eq:E}) and (\ref{eq:Y}), the following relation is constructed,
\begin{equation}
\begin{aligned}
&\beta\left(dE-Ydy+\frac{\alpha}{\beta}dN\right)\\
=&-\beta d\left(\frac{\partial \ln_qZ_q}{\partial\beta}\right)+\frac{\partial\ln_qZ_q}{\partial y}dy-\alpha d\left(\frac{\partial\ln_qZ_q}{\partial\alpha}\right).
\end{aligned}
\label{eq:constructed_id_1}
\end{equation}
Since $\ln_qZ_q$ is a function of $\alpha$, $\beta$, and $y$, its full differential is
\begin{equation}
d(\ln_qZ_q)=\frac{\partial \ln_qZ_q}{\partial\beta}d\beta+\frac{\partial\ln_qZ_q}{\partial\alpha}d\alpha+\frac{\partial\ln_qZ_q}{\partial y}dy.
\label{eq:constructed_id_2}
\end{equation}
Combing Eqs. (\ref{eq:constructed_id_1}) and (\ref{eq:constructed_id_2}), we can get
\begin{equation}
\begin{aligned}
&\beta\left(dE-Ydy+\frac{\alpha}{\beta}dN\right)\\
=&d\left(\ln_qZ_q-\alpha\frac{\partial \ln_qZ_q}{\partial\alpha}-\beta\frac{\partial \ln_qZ_q}{\partial\beta}\right).
\end{aligned}
\label{eq:constructed_id_3}
\end{equation}
It has been shown that the following thermodynamic relation still holds in the non-extensive statistics \cite{nex18,non_ext_hydro}
\begin{equation}
dS=\frac{1}{T}(dE-Ydy-\mu dN).
\label{eq:ther_relation}
\end{equation}
Comparing Eq. (\ref{eq:constructed_id_3}) with Eq. (\ref{eq:ther_relation}), we obtain the entropy $S$ as
\begin{equation}
\begin{aligned}
S=\ln_qZ_q-\alpha\frac{\partial \ln_qZ_q}{\partial\alpha}-\beta\frac{\partial \ln_qZ_q}{\partial\beta}.
\end{aligned}
\label{eqA14}
\end{equation}
Therefore, in the non-extensive statistics the relation between the grand-canonical potential $\Omega$ and the grand-canonical partition function $Z_q$ is
\begin{equation}
\begin{aligned}
\Omega=&E-TS-\mu N\\
=&-\frac{\partial}{\partial\beta}\ln_qZ_q-\mu\left(-\frac{\partial}{\partial\alpha}\ln_qZ_q\right)\\
&-T\left(\ln_qZ_q-\alpha\frac{\partial \ln_qZ_q}{\partial\alpha}-\beta\frac{\partial \ln_qZ_q}{\partial\beta}\right)\\
=&-T\ln_qZ_q.
\end{aligned}
\label{eqA15}
\end{equation}

\newpage
%\vspace{10.0cm}

%\bibliography{apssamp}% Produces the bibliography via BibTeX.

\end{document}